\documentclass{ws-procs9x6-cpt16}
\usepackage{graphicx}

\begin{document}

\newcommand{\refeq}[1]{(\ref{#1})}
\def\etal {{\it et al.}}

\title{Lorentz Violation in Deep Inelastic Electron-Proton Scattering}

\author{E.\ Lunghi}

\address{Physics Department, Indiana University, Bloomington, IN 47405, USA}

\begin{abstract}
Lorentz violation in the quark sector induces a sidereal time dependence in electron-proton, proton-antiproton and proton-proton cross sections. At high energies nonperturbative effects are buried in universal nucleon parton distribution functions and Lorentz violating effects are calculable in perturbation theory. We focus on deep inelastic electron-proton scattering data collected from ZEUS and H1 at HERA and show that a sideral time analysis of these events is able to set strong constraints on most of the coefficients we consider.
\end{abstract}

\bodymatter

\phantom{}\vskip10pt\noindent
We consider Lorentz violation in the quark sector as parametrized by the following lagrangian:\cite{CK1}
\begin{align}
{\cal L}_{\rm quark}=\frac{1}{2}i(g^{\mu\nu}+ c^{\mu\nu}_f) (\bar{\psi}\gamma_{\mu}\overleftrightarrow{D}_{\nu}\psi+2iQ_f\bar{\psi}\gamma_{\mu}A_{\nu}\psi)\; ,
\label{eq1}
\end{align}
where the coefficients $c^{\mu\nu}_f$ ($f=u, d$) are poorly constrained.\cite{tables} These coefficients are obviously related to the corresponding coefficients for proton and neutron but the connection is heavily clouded by nonperturbative physics.

In a recent work\cite{paper} we study the effects of these coefficients on the deep inelastic electron-proton cross section for which factorization theorems show that nonperturbative effects are confined to universal nucleon parton distribution functions. Details of the calculation can be found in Ref.\ \refcite{paper,vieira}. Here we present a short summary of the main phenomenological results. 

We focus on the complete set of deep inelastic $e p \to e X$ measurements performed by the ZEUS and H1 experiments at the HERA collider.\cite{Abramowicz:2015mha} Note that these measurements enter the global fits used to extract the proton parton distribution functions, therefore it is impossible to extract bounds on the coefficients $c^{AB}_f$ from measurements of the sidereal time integrated cross section.

\begin{table}[b]
\tbl{Expected 95\% C.L. upper limits on the magnitude of the coefficients $c_f^{XZ}$, $c_f^{XY}$, $c_f^{YZ}$, $c_f^{TX}$, $c_f^{TY}$ and $c_f^{XX} - c_f^{YY}$. The second column shows the best expected upper limit coming from a single HERA measurement. The third column the upper limit extracted from a single sidereal time analysis of the whole HERA data set. \label{eltab:results}}
{
\begin{tabular}{|c|c|c|} \hline
& Best individual  & Overall limit \vphantom{$\Big($}\cr 
& limit on $|c_{AB}^f|$ &  on $|c_{AB}^f|$ \vphantom{$\Big($}\cr \hline\hline
$|c^{XZ}_u|$ & $4.1 \times 10^{-5}$ & $4.6 \times 10^{-6}$  \phantom{$\Big($}\cr\hline
$|c^{XY}_u|$ & $2.9 \times 10^{-5}$ & $2.3 \times 10^{-6}$  \phantom{$\Big($}\cr\hline
$|c^{YZ}_u|$ & $4.0 \times 10^{-5}$ & $4.8 \times 10^{-6}$   \phantom{$\Big($}\cr\hline
$|c^{TX}_u|$ & $3.3 \times 10^{-5}$ & $9.3 \times 10^{-6}$  \phantom{$\Big($}\cr\hline
$|c^{TY}_u|$ & $3.3 \times 10^{-5}$ & $9.1 \times 10^{-6}$  \phantom{$\Big($}\cr\hline
$|c^{XX}_u - c^{YY}_u|$ & $1.7 \times 10^{-5}$ & $7.1 \times 10^{-6}$  \phantom{$\Big($}\cr\hline\hline
$|c^{XZ}_d|$ & $3.5 \times 10^{-5}$ & $1.9 \times 10^{-5}$ \phantom{$\Big($}\cr\hline
$|c^{XY}_d|$ & $1.7 \times 10^{-5}$ & $8.8 \times 10^{-6}$  \phantom{$\Big($}\cr\hline
$|c^{YZ}_d|$ & $3.5 \times 10^{-5}$ & $1.9 \times 10^{-5}$  \phantom{$\Big($}\cr\hline
$|c^{TX}_d|$ & $1.2 \times 10^{-4}$ & $7.1 \times 10^{-5}$  \phantom{$\Big($}\cr\hline
$|c^{TY}_d|$ & $1.2 \times 10^{-4}$ & $6.8  \times 10^{-5}$  \phantom{$\Big($}\cr\hline
$|c^{XX}_d - c^{TT}_d|$ & $4.8 \times 10^{-5}$ & $2.8 \times 10^{-5}$  \phantom{$\Big($}\cr\hline
\end{tabular}
}
\end{table}

After taking into account the location and orientation of the ZEUS and H1 detectors the relation between the coefficients $c^{\mu\nu}_f$ in the laboratory frame and the Sun-centered frame acquires a sidereal time ($t$) dependence. The cross section at fixed values of Bjorken $x$ and $Q^2$ (electron 4-momentum transfer squared) can be schematically written as:
\begin{align}
\sigma (t,x,Q^2) &= \sigma_{\rm SM} (x, Q^2) \; \Bigg[
 1 +
\sum_{\substack{AB=XX+YY,\\ZZ,TT,ZT}} c^{AB}_f \; \alpha_{AB}^f   \nonumber \\
&+ \sum_{\substack{AB=XZ,YZ,\\TX,TY}} \left(c^{AB}_f \; \beta_{AB}^f  \cos \Omega t 
 +  c^{AB}_f  \;\gamma_{AB}^f  \sin \Omega t \right)\nonumber \\
& + \sum_{\substack{AB=XY,\\XX-YY}} \left( c^{AB}_f  \;\delta_{AB}^f  \cos 2\Omega t 
 + c^{AB}_f \; \epsilon_{AB}^f \sin 2\Omega t \right)
\Bigg] \; ,
\label{eq:xs}
\end{align}
where $\Omega$ is the sidereal time frequency and $(\alpha,\beta,\gamma,\delta,\epsilon)_{AB}^f$ are functions of $x$ and $Q^2$. These functions are completely determined in terms of fundamental Standard-Model parameters (that we take from the Particle Data Group\cite{Agashe:2014kda}) and of the parton distribution functions of the proton (for our central values we adopt the CT10 PDFs set\cite{Lai:2010vv} and use the program ManeParse\cite{Godat:2015xqa, Clark:2016jgm}). Note that the coefficients $c_f^{ZZ,TT,ZT}$ enter without any sidereal time dependence and cannot be constrained by our analysis.

\begin{figure}[b] 
\centering
\includegraphics[width=0.52 \linewidth]{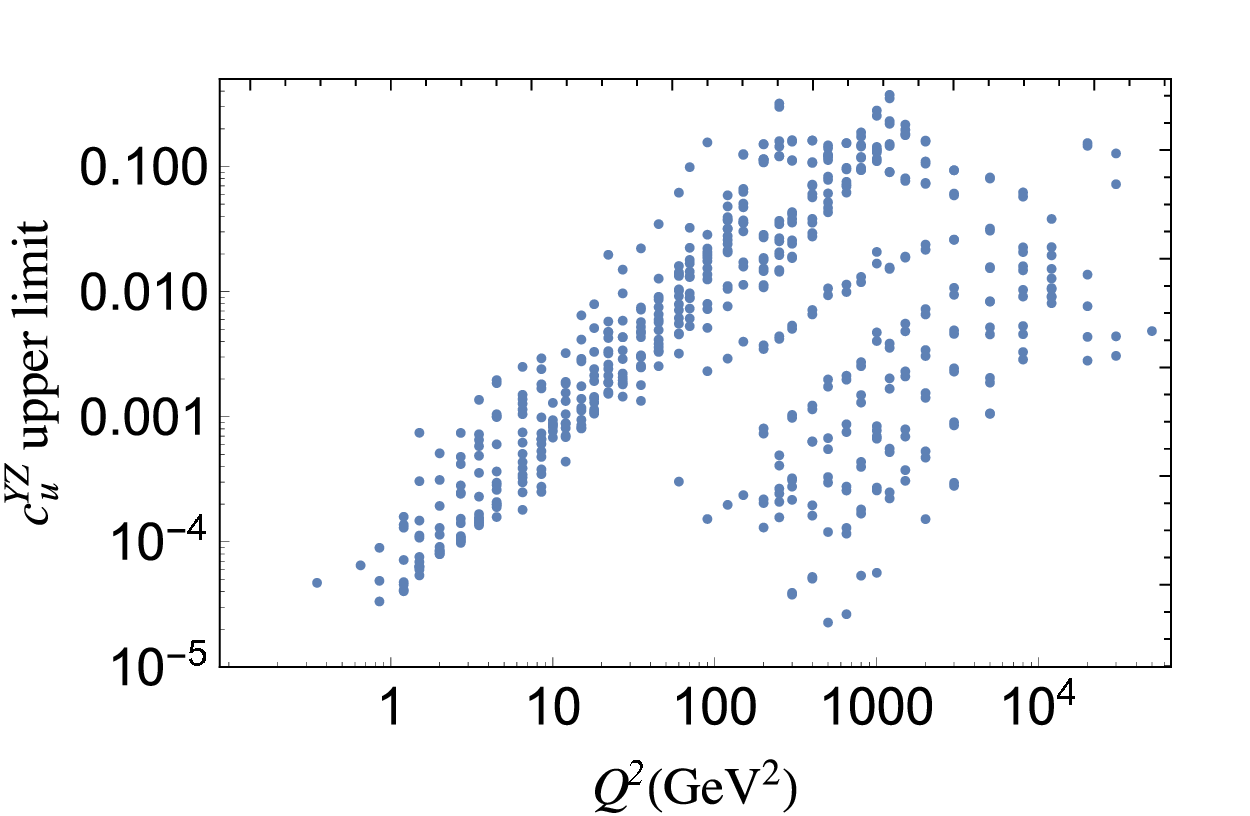}
\includegraphics[trim = 0mm -3mm 0mm 0mm, width=0.433 \linewidth]{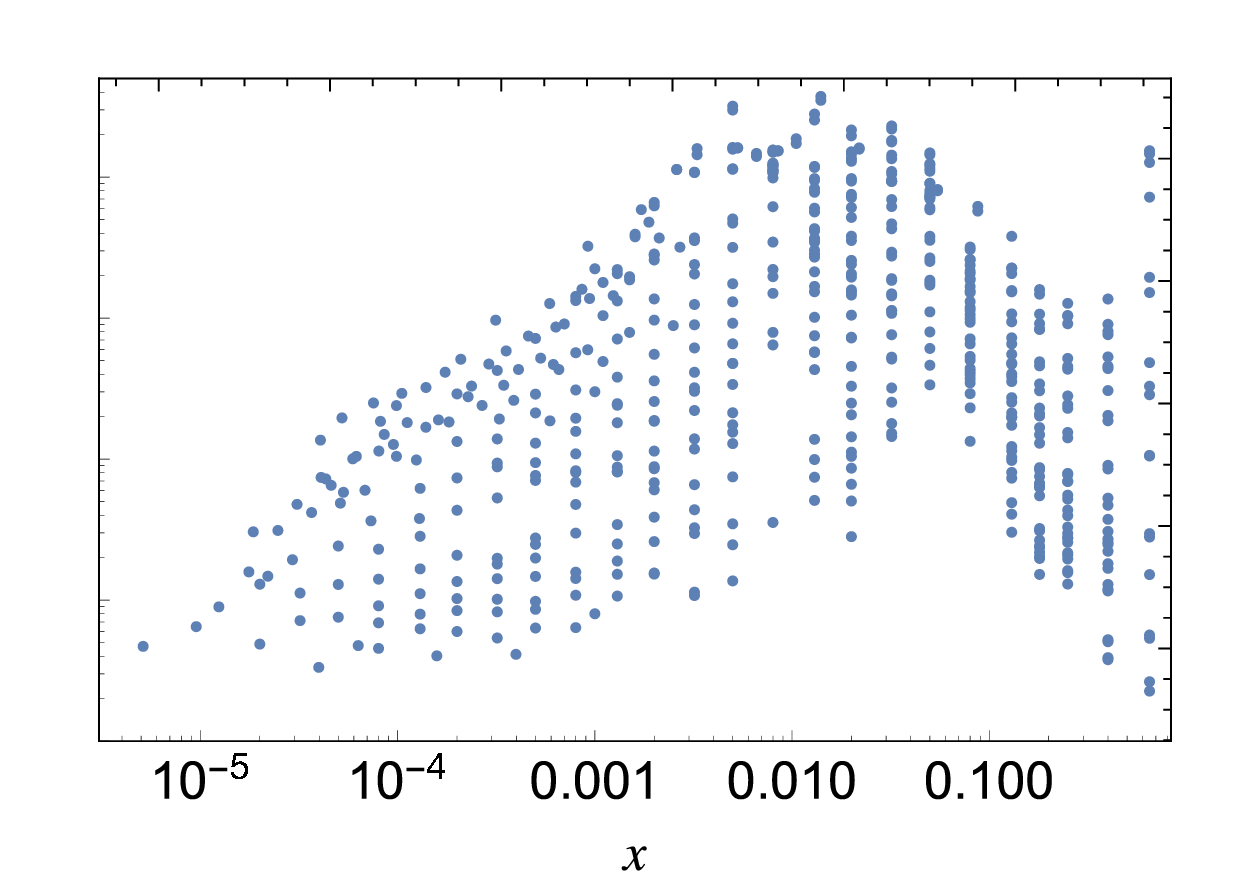}
\includegraphics[width=0.52 \linewidth]{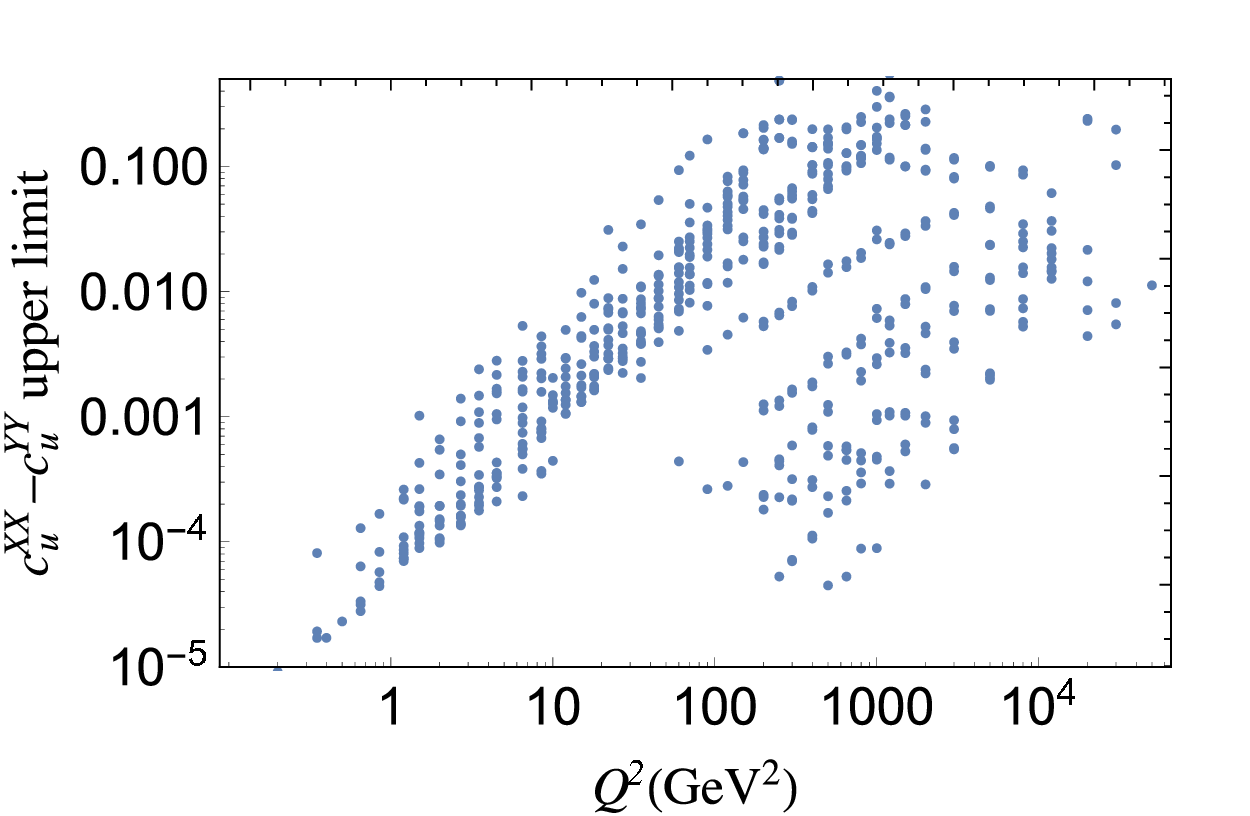}
\includegraphics[trim = 0mm -3mm 0mm 0mm, width=0.433 \linewidth]{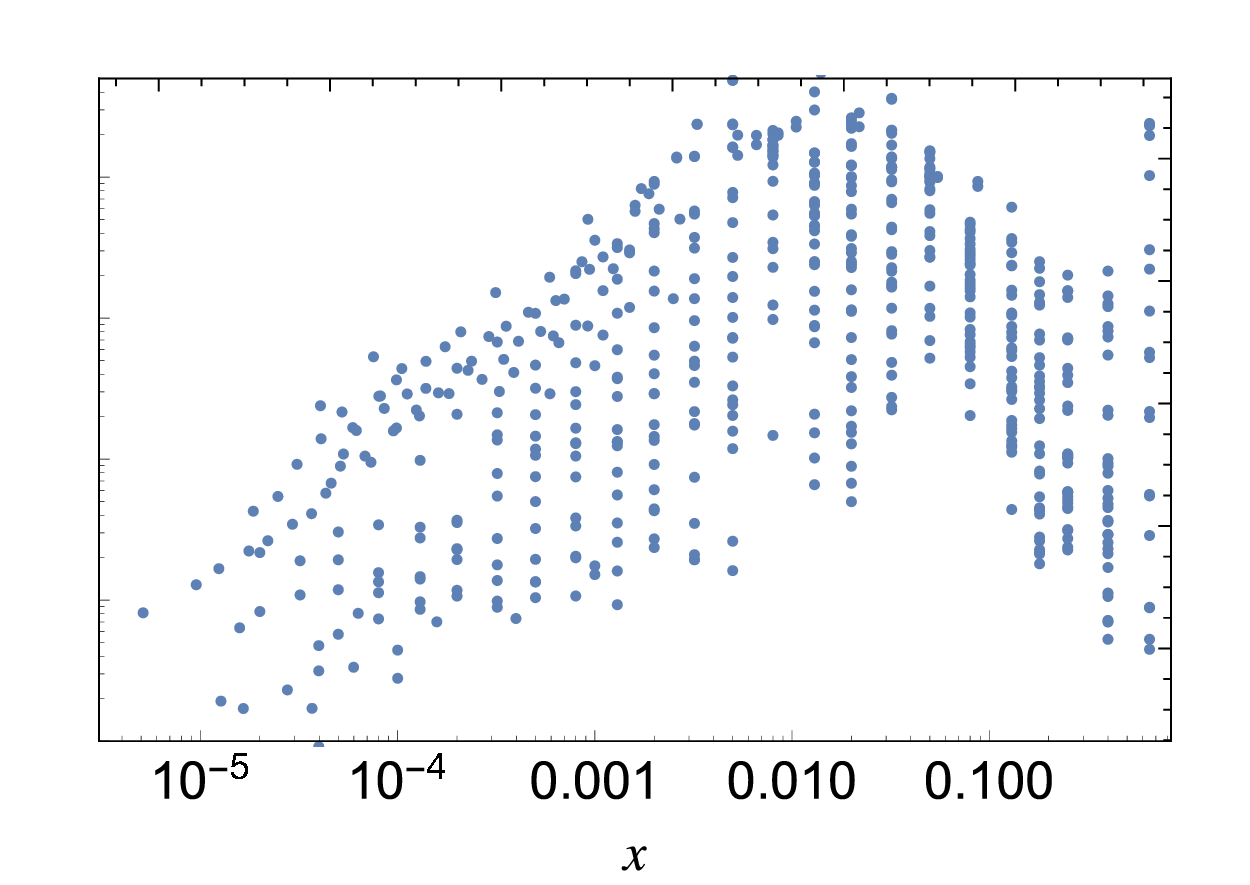}
\caption{Expected 95\% C.L. upper limits on the magnitude of the coefficients $c_u^{YZ}$, $c_u^{XX}-c_u^{YY}$ from a 4-bin analysis of HERA data.
Each point corresponds to a measurement with fixed $x$ and $Q^2$. \label{fig_bounds_up}}
\end{figure}

For each HERA measurement we integrate Eq.~(\ref{eq:xs}) in four sidereal time bins and set to zero all but one of the $c^{AB}_f$ coefficients. The expected upper bounds that this analysis might imply can be calculated by generating a large number of pseudoexperiments that are consistent with the sidereal time averaged results (i.e., the published cross sections). For each pseudoexperiment we write a chi squared and extract the 95\% upper limit on $|c^{AB}_f|$. The expected upper limit is the median of this quantity over all pseudoexperiments. Since the orientation of the ZEUS and H1 experiments are opposite to each other the rotations required to connect the laboratory frames at ZEUS and H1 to the Sun-centered frame are different, but this has no effect on the bounds for individual coefficients $c^{AB}_f$.

The results of this analysis are presented in the first column of Table~\ref{eltab:results} where we consider the 12 coefficients $c_f^{XZ}$, $c_f^{XY}$, $c_f^{YZ}$, $c_f^{TX}$, $c_f^{TY}$, and $c_f^{XX}-c_f^{YY}$ ($f=u,d$) and show the constraints that a potential sidereal time analysis of ZEUS and H1 data might produce. Here we consider each HERA measurement (at fixed $x$ and $Q^2$) and show the strongest constraint. In Fig.~\ref{fig_bounds_up} we show the upper limits expected from a sidereal time analysis of HERA data on $c^{YZ}_u$ and $c^{XX}_u-c^{YY}_u$ as a function of  $Q^2$ (left plot) and $x$ (right plot); each point corresponds to one of the 644 neutral current HERA measurements.\cite{Abramowicz:2015mha} The strongest constraints come mostly from measurements at low $Q^2$ and low $x$ (very close to the kinematical boundary $Q^2 = s x$). Constraints on these two coefficients are representative of constraints on $c_{ZX,XY,ZY,TX,TY}^{u,d}$ and $c_{XX-YY}^{u,d}$. 

Finally we perform a global sidereal time analysis of the whole HERA data set by combining all measurements into a single chi squared. The expected upper limits that we obtain are listed in the third column of Table~\ref{eltab:results}. These constraints are stronger than the best single measurement limits because this analysis makes full use of the correlations between the binned integrated cross sections at different values of $x$ and $Q^2$. 

In conclusion, we found that a sidereal time analysis of ZEUS and H1 data has the potential to set strong constraints on most $c^{AB}_f$ coefficients.

\section*{Acknowledgments}
This work was supported in part by the Indiana University
Center for Spacetime Symmetries.


\begin{thebibliography}{x}

\bibitem {CK1}
D.\ Colladay and V.A.\ Kosteleck\'y,
Phys.\ Rev.\ D {\bf 58}, 116002 (1998).

\bibitem{tables}
{\it Data Tables for Lorentz and CPT Violation,}
V.A.\ Kosteleck\'y and N.\ Russell,
2016 edition,
arXiv:0801.0287v9.

\bibitem{paper}
V.A.\ Kosteleck\'y, E.\ Lunghi, and A.R.\ Vieira, 
arXiv:1610.08755.

\bibitem{vieira} 
A.R.\ Vieira, these proceedings.

\bibitem{Abramowicz:2015mha} 
H.\ Abramowicz \etal,
Eur.\ Phys.\ J.\ C {\bf 75}, 580 (2015).

\bibitem{Agashe:2014kda} 
K.A.\ Olive \etal,
Chin.\ Phys.\ C {\bf 38}, 090001 (2014).

\bibitem{Lai:2010vv} 
H.L.\ Lai, M.\ Guzzi, J.\ Huston, Z.\ Li, 
P.M.\ Nadolsky, J.\ Pumplin, and C.-P.\ Yuan,
Phys.\ Rev.\ D {\bf 82}, 074024 (2010).

\bibitem{Godat:2015xqa} 
E.\ Godat,
arXiv:1510.06009.

\bibitem{Clark:2016jgm} 
D.B.\ Clark, E.\ Godat and F.I.\ Olness,
arXiv:1605.08012.


\end{thebibliography}
\end{document}